\documentclass[final,5p,times,twocolumn]{elsarticle}

\usepackage{graphics,graphicx,dcolumn,bm,fleqn,epic,eepic,float,ulem}
\usepackage{amssymb,amsmath}
\usepackage{color}

\definecolor{red}{rgb}{1,0,0}
\definecolor{green}{rgb}{0,1,0}
\definecolor{blue}{rgb}{0,0,1}

\journal{Physics Letters A}
\begin{document}
%%%%%%%%%%%%%%%%%%%%%%%%%%%%%%%%%%%%%%%%%%%%%%%%%%%%%%%%%%%%%%%%%%%%%%%%%

\begin{frontmatter}

\title{The bounds of heavy-tailed return distributions in evolving
complex networks}
%to measure model risk}

\author{Jo\~ao P.~da~Cruz$^{a,b,c}$}
\author{Pedro G.~Lind$^{b,{\ast}}$}

\address{$^a$Closer, Consultoria, Lda., 
         Av.~Eng.~Duarte Pacheco, Torre 2, 14º~C
         1070-102 Lisboa, Portugal}
\address{$^b$Center for Theoretical and Computational Physics, 
         University of Lisbon, Av.~Prof.~Gama Pinto 2, 
         1649-003 Lisbon, Portugal}
\address{$^c$Departamento de F\'{\i}sica, Faculdade de Ci\^encias 
         da Universidade de Lisboa, 1649-003 Lisboa, Portugal} 
\address{$^{\ast}$ Phone: +351 21 790 4862; Fax: +351 21 795 4288; Email: plind@cii.fc.ul.pt}

\begin{abstract}
We consider the evolution of scale-free
networks according to preferential attachment schemes and show 
the conditions for which the exponent characterizing the degree
distribution is bounded by upper and lower values.
Our framework is an agent model, presented in the context of
economic networks of trades, which shows the emergence of critical behavior.
Starting from a brief discussion about the main features of the evolving 
network of trades, we show that the logarithmic return distributions have 
bounded heavy-tails, and the corresponding  
bounding exponent values can be derived.
Finally, we discuss these findings in the context of model risk.
\end{abstract}

%%%%PACS e Keywords
\begin{keyword}
Agent-based model \sep
Criticality and Crisis \sep
Model risk \sep
%Predictability
\PACS[2010] 05.65.+b  \sep   %Self-organized criticality
      05.40.-a \sep   %Stochastic processes 
      89.65.Gh \sep   %Econophysics
      89.65.Gh \sep   %Economics: business and management

\end{keyword}

\end{frontmatter}

%%%%%%% Table of contents
\tableofcontents

% \linenumbers

%%%%%%%%%%%%%%%%%%%%%%%%%%%%%%%%%%%%%%%%%%%%%%%%%%%%%%%%%%%%%%%%%%%%%%%%%
%%%%%%%%%%%%%%%%%%%%%%%%%%%%%% TEXT %%%%%%%%%%%%%%%%%%%%%%%%%%%%%%%%%%%%%
%%%%%%%%%%%%%%%%%%%%%%%%%%%%%%%%%%%%%%%%%%%%%%%%%%%%%%%%%%%%%%%%%%%%%%%%%

%%%%%%%%%%%
\section{Introduction: a note on agent-models for social systems}
\label{intro}

Similarly to other fields in social sciences, 
most of the research made in finance and economics has been 
dominated by an epistemological approach, in which the behavior of the 
economic system is explained by a few key characteristics of the behavior 
itself, like the amplitude of price fluctuations or the analytical form of 
the heavy-tailed return 
distributions \cite{Mantegna_Stanley1999,Borland_Bouchaud2005}. 
These key characteristics motivated researchers to assume
such distributions 
as $\alpha$-stable L\' evy distributions 
or truncated $\alpha$-stable L\' evy  
distributions \cite{Mantegna_Stanley1994}.
The reason for this assumption is given by the more general
version of the central limit theorem -- sometimes not so well known --
which states that the aggregation of a growing number of random variables 
converges to a $\alpha$-stable L\'evy distribution\cite{Kolmogorov54}. 
If these random variables have finite variances then the resulting 
aggregation is a $2$-stable L\'evy distribution, 
i.e.~a Gaussian distribution. 
If the variances are infinite -- or of the order of the system size -- 
then $\alpha<2$ and the so-called heavy-tailed shape emerges as a 
result of the aggregation.
Further, non-Gaussian (heavy-tailed) distributions are associated with
correlated variables and therefore it is reasonable to 
assume that measurements on aggregates of human activities will result in 
a $\alpha-$stable L\'evy distribution, since humans are strongly 
correlated with each other.
Henceforth, we refer to $\alpha$-stable L\' evy distributions with 
$\alpha <2$ as L\' evy distributions and with $\alpha =2$  
as Gaussian distributions. 
  
Without leaving an epistemological approach, we could address the study of the 
resulting distributions by ignoring 
the previous arguments and construct a function 
that fits any set of empirical data
just by building up fitting parameters until the plotted function 
fit the empirical data. 
Such approach would be the best one, if economic processes were stationary.
Unfortunately they are not \cite{Bouchaud,Sornette,CruzLind1} 
and this means that we cannot disregard the underlying mechanisms 
generating the data we are analyzing.

Since heavy-tails are observed in the returns of economic
variables, one would expect that practitioners use L\'evy distributions. 
The particular case of Gaussian distribution was the first to
be considered for modeling price of European options, through the well 
known Black-Scholes model\cite{Black} proposed in 1973.
This model ended a story started already in 1900 with Bachelier and his 
{\it Theory of Speculation}\cite{Bachelier} where Brownian motion was
used to model stock price evolution.
The Black-Scholes model for option-pricing is however inconsistent with 
options data, since stock-price behavior is essentially not Gaussian.
To overcome the imperfections of the Black-Scholes model, more sophisticated
models were proposed since 1980s and 1990s, which basically assume 
processes more general than Brownian processes.
These processes are called L\'evy processes\cite{Schoutens} and the 
probability distributions of their increments are infinitely divisible, 
i.e.~one random variable following that probability distribution can be
decomposed into one sum of an arbitrary integer number of independent 
identically distributed random variables.

Still, despite considerable progresses on modeling financial data with 
L\'evy processes, practitioners continue to show a strong preference for 
the particular class of finite moment's distributions and there are good 
reasons for that. 
Assuming that L\'evy distributions are good representations of economic 
variables fluctuations, a model based on them is closed when one fits the 
distribution to empirical data choosing properly the parameter values,
which represent the valuable information for financial insight and
decision making.
However, as said above, 
fitting is no good when the series are not stationary:
There is no guarantee that today's fitting will be the same as tomorrows.
%unlike the Gaussian distribution that can characterize an infinity of 
%systems, a heavy-tailed Lévy distribution constitutes a signature of the 
%state of the underlying system which means that empirical fitting is, in 
%general, a wrong procedure if the stationarity of the system is not 
%guaranteed.
Since working with a Gaussian curve is more straightforward
than working with a  L\'evy distribution and needs less parameters for curve 
fitting, there is no practical gain in abandoning Gaussian distribution 
to model the distribution of fluctuations according to a prescribed 
mathematical model, even though it is not entirely correct. 
In other words, if a L\'evy distribution is fitted to empirical 
data of a non-stationary process one will carry basically the same 
{\it model risk}, as if a Gaussian distribution is used.

On a more ontological approach, when modeling financial 
and economic networks, random variables 
are translated into agents. Agent-based models for describing and
addressing the evolution of markets has become an issue of
increasing interest \cite{abmfarmer} and appeals for further 
developments \cite{yakovenko09,HaldaneMay,JohnsonLux,LuxStaufferReview}.
They enable one to access three important questions\cite{Bouchaud}.
First, the system is able in this way to be decomposed into sellers
and buyers, a common feature of all finance systems. 
Second, one enables non-stationary regimes to occur, as in real stock 
markets. 
Third, by properly incorporating the ingredients of financial agents 
and the trades among them one can directly investigate the impact
of trades in the price, according to some prescribe scheme.

In this paper we use an agent model for the individual behavior 
of single financial agents, at a microscopic scale, in a way that 
the collective behavior generates an output in accordance with the 
observed curves of macroscopic variables, namely the financial indices. 
Several of such bottom-up approaches were thoroughly 
investigated\cite{Lux_Marchesi,yakovenko09}.
The Solomon-Levy model \cite{Solomon_Levy} defines each agent as a 
wealth function $\omega_i(t)$ that cannot go below a floor level, 
given by $ \omega_{i}(t) \geq \omega_{0}\bar{\omega}(t)$ where 
$\bar{\omega}(t)$ is the agent average $\omega$ at instant $t$
and $\omega_0$ is a proper constant. 
The imposition of the floor based on the mean field $\bar{\omega}(t)$ 
means that on average $\langle \left| \omega_i(t) - 
\bar{\omega}(t) \right|\rangle 
\sim N $ and, by basic 
statistics, $\mathrm{var}(\omega(t)) \sim N^2$. 
Consequently, the result of the Solomon-Levy model, despite the interesting 
idea of the introduction of a floor similar to what was done by 
Merton\cite{Merton} in the agent dynamics, will surely be a $\alpha$-stable 
distribution with a power law heavy-tail, i.e.~$\alpha<2$.
%{\it A fortiori}, 
Percolation based models like 
Cont-Bouchaud \cite{Cont_Bouchaud} or Solomon-Weisbuch \cite{Solomon_Weisbush}, 
by the nature of the phenomena, also brings up variations of the order 
of the system size, leading also to L\'evy-type distributions. 

In our approach, we follow the above considerations, 
to address the following question: what are the fundamental assumptions, 
common e.g.~to all economic systems, that naturally lead to the emergence 
of macroscopic distributions that are characterized by heavy-tails? 
Taking an economic system as a prototypical example for the emergence of
heavy-tailed distributions, we argue that there are three fundamental 
assumptions.

First, agents tend to trade, i.e.~to interact. 
Human beings are more efficient in doing 
specialized labor than being self-sufficient and for that they need to 
exchange labor. 
The usage of the expression `labor' can be regarded as 
excessive by economists, but we look at it as the fundamental 
quantity that is common to labor, money or wage.
Something must be common to all these quantities; if not,
we wouldn't exchange them. The physicists can regard such fundamental
quantity as an `economic energy'. 

Second, we only consume and produce a finite amount of the overall 
product that exists within our environment.
This assumption justifies the emergence for each agent of a maximum 
production and minimum consumption.
If an agent transposes that finite amount he should not be able to
consume anymore. 

Third, human agents are different and attract differently other 
agents to trade. 
For choosing the way ``how'' agents attract each other for trading,
we notice that this heterogeneity should reflect some imitation, where
agents tend to prefer to consume (resp.~produce) from (resp.~to)
the agents with the largest number of consumers (resp.~producers).
The number of producer and consumer neighbors
reflects, respectively, supply and demand of its labor.
With such observation its is reasonable to assume that
combining both kinds of neighbors should suffice to 
quantify the price of the labor exchanged.

Heavy-tailed distributions have been subject to intensive research
activity till very
recently, e.g.~when addressing the formation and construction of
efficient reservoir networks\cite{nuno}, which shows self-organized
criticality with critical exponents that can be explained by
a self-organized-criticality-type model.
In this paper, we deal with heavy tails found in economic systems and
show that heavy-tailed return distributions are due to 
the economic organization emerging in a complex economic network of 
trades among agents governed under the above three assumptions.
Further, the model reproducing empirical data is also of the
self-organized-criticality-type model, but its main ingredients
result from economical reasoning and assumptions.

Our central result deals in particular with the return distribution
found in both data and model:
we show that the power-law tails are characterized by an exponent
that can be measured and is constrained by upper and lower bounds, which
can be analytically deduced. 
The knowledge of such boundaries is of great importance for 
risk estimates: by deriving upper and lower bounds, one avoids either
underestimates, which enable the occurrence of crisis unexpectedly, 
as well as overestimates, which prevent profit maximization of the 
trading agents.

We start in Section 
%\ref{sec:topology} by discussing the emergence of
%heavy-tailed distributions in the topology underlying the economic 
%network.
%In section 
\ref{sec:criticality} by describing the ubiquity of heavy-tails
in financial time-series, namely in stock indices.
We will argue that such heavy-tails result from the combination
of a dynamical critical state in real economic systems and an
underlying scale-free topology.
%we will describe 
%the model that reproduces
%the critical behavior observed in real economic systems and which
%leads to a critical distribution, typically heavy-tailed.
%The critical state combined with the underlying topology leads to 
%a limited range of admissible exponent values describing the heavy-tails. 
Applied to a real system such as the financial market,
such bounded behavior leads naturally to a maximum and minimum value 
on risk evaluation, improving the knowledge about the uncertainty of 
the market future evolution. 
These bounding values will be derived in Sec.~\ref{distributionbounderies}   
based in the assumptions listed above and an application to risk model
is discussed.
Section \ref{conclusion} concludes this paper.
%%%%%%%%%%%%%%%%%%%%%%%%%%%%%%%%%%%%%%%%%%%%%%%%%%%%%%%%%%%%%%%%%%%%%%%%%%
\begin{figure}[htb]
\centerline{\includegraphics[width=0.99\linewidth]{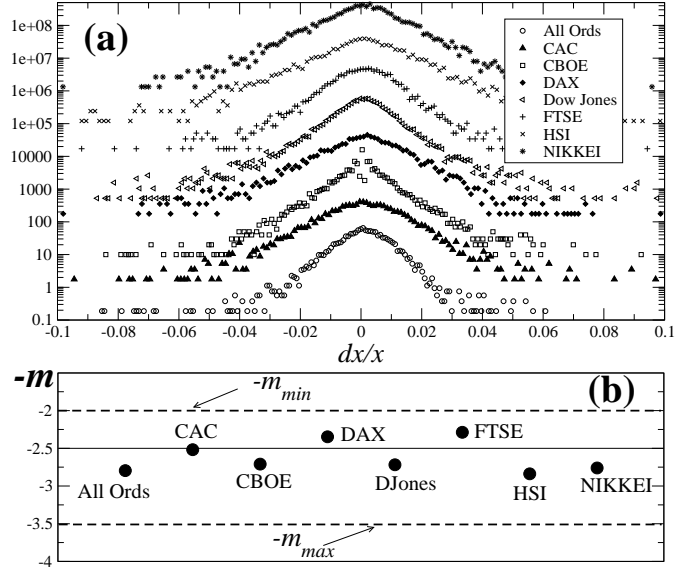}}
\vspace*{8pt}
\caption{\protect
         {\bf (a)} Heavy-tails of stock indices cumulative distributions 
         and {\bf (b)} the corresponding exponents.
         Distributions in (a) are shifted in the vertical axis for easy 
         comparison.
         Solid line in (b) ($m\sim 5/2$) indicates the value obtained 
         with the numerical model introduced in Ref.~\cite{CruzLind1}, 
         while dashed
         lines indicate the bounding values of $m$ for $2 < \gamma < 3$
         (see Eq.~(\ref{m})).}
\label{fig3} 
\end{figure}
%%%%%%%%%%%%%%%%%%%%%%%%%%%%%%%%%%%%%%%%%%%%%%%%%%%%%%%%%%%%%%%%%%%%%%%%%%

%%%%%%%%%%%%%%%%%%%%%%%%%%%%%%%%%%%%%%%%%%%%%%%%%%%%%%%%%%%%%%%%%%%%%%%%%%%
\section{Critical behavior underlying return distributions}
\label{sec:criticality}

Heavy-tails are observed in return distributions of data in finance and
economics.
Figure \ref{fig3} presents data from several stock market indices.
Figure \ref{fig3}a shows the probability density functions (PDF)
of the logarithmic returns of each index, symbolized as $x$, 
where one can observe the heavy-tails.
The exponent characterizing the tails of these distributions
are given in Fig.~\ref{fig3}b.

While the heavy-tailed shape of the return distributions was already 
known and several times reported\cite{Mandelbrot2}, 
%at least within a good approximation, 
the explanation for their emergence,
and in particular the values of the exponent characterizing them,
was up to our knowledge not so frequently addressed.
%addressed till now.}
%%%%%%%%%%%%%%%%%%%%%%%%%%%%%%%%%%%%%%%%%%%%%%%%%%%%%%%%%%%%%%%%%%%%%%%%%%
\begin{figure}[htb]
\centerline{\includegraphics[width=0.5\textwidth]{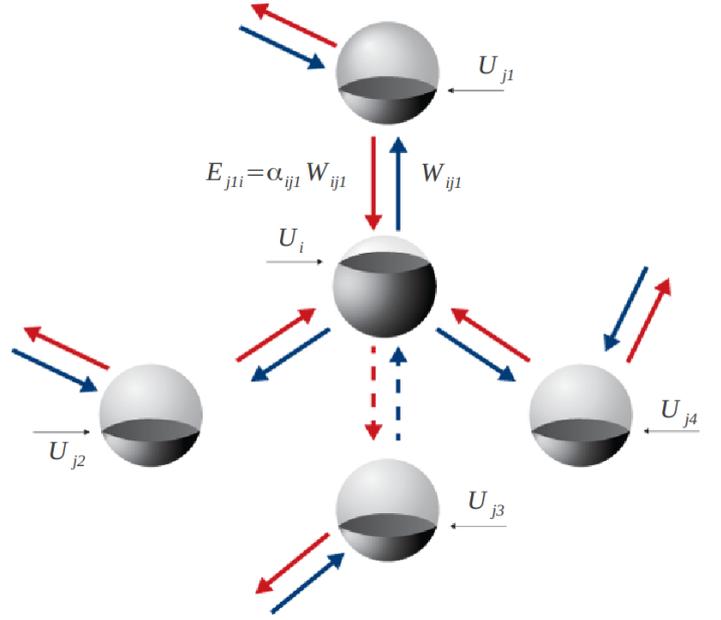}}
\vspace*{8pt}
\caption{\protect
  Illustration of economical connections between economic agents. 
  Agent $i$ transfers an amount of ``labor'' $W_{ij_1}$ to agent $j_1$ 
  receiving in return an amount $E_{ij_1}=\alpha_{ij_1}W_{ij_1}$ where 
  $\alpha_{ij_1}$ measures how well the labor is rewarded.
  For this trade ``interaction'' agent $i$ establishes an outgoing 
  connection with (production to) agent $j_1$, while agent $j_1$ establishes 
  an incoming connection with (consumption from) agent $i$.
  The balance of this interaction yields for agent $i$ an amount
  of ``internal energy'' $U_i=\sum_{j_k} U_{ij_k}=\sum_{j_k} (W_{ij_k}-E_{ij_k})$ 
  that can be summed up over all agents connections. }
\label{fig1}
\end{figure}
%%%%%%%%%%%%%%%%%%%%%%%%%%%%%%%%%%%%%%%%%%%%%%%%%%%%%%%%%%%%%%%%%%%%%%%%%%

The emergence of the heavy-tails of the return distribution was
recently reproduced with a simple model\cite{CruzLind1} which takes
one economic connection as an exchange of labor between two agents,
say $i$ and $j$, dissipating an amount of energy $U_{ij}$, representing
the deficit of $i$ that results from the labor exchange between $i$ 
and $j$. 
Agent $i$ delivers an amount of labor $W_{ij}$ to agent $j$ and gets a 
proportional amount of ``reward'' $E_{ij}=\alpha_{ij} W_{ij}$
where $\alpha_{ij}$ can be interpreted as a `exchange rate' of labor. 
Figure \ref{fig1} illustrates the economic connection between two 
agents.
%Being directed from $i$ to $j$, agent $i$ takes the connection as
%a production (outgoing) connection and agent $j$ as a consumption  
%(incoming) connection.
Each agent can connect to several others,
since every agent will have the propensity to establish new economic 
links to get specialized labor from other agents.
Consequently the ``internal economic energy'' $U_i$ at one single agent
will sum up all differences between labor units $W_{ij}$ and the
corresponding reward $E_{ij}$ as explained in Fig.~\ref{fig1}, weighted
by the coefficient $\alpha_{ij}$.
This coefficient takes into account how many
incoming and outgoing connections the agent and its neighbors have, 
i.e.~how much demand and supply they have respectively\cite{CruzLind1}.
\begin{figure}[htb]
\centerline{\includegraphics[width=0.49\textwidth]{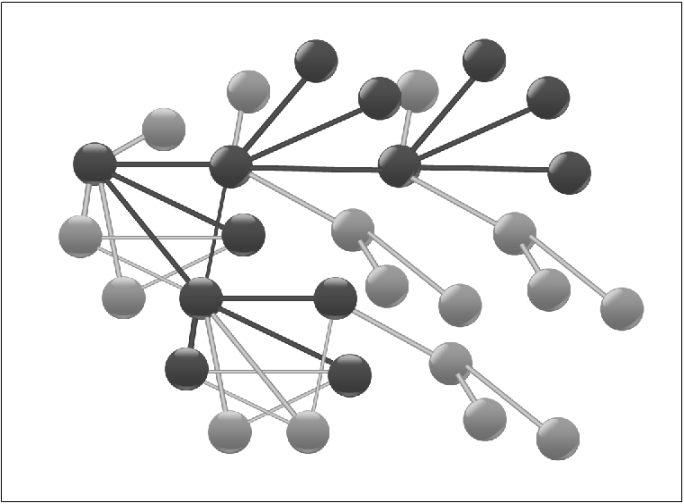}}
\vspace*{8pt}
\caption{\protect
         Illustration of a collapse chain reaction.}
\label{fig2}
\end{figure}
%%%%%%%%%%%%%%%%%%%%%%%%%%%%%%%%%%%%%%%%%%%%%%%%%%%%%%%%%%%%%%%%%%%
%%%%%%%%%%%%%%%%%%%%%%%%%%%%%%%%%%%%%%%%%%%%%%%%%%%%%%%%%%%%%%%%%%%
\begin{figure}[htb]
\centerline{\includegraphics[width=0.49\textwidth]{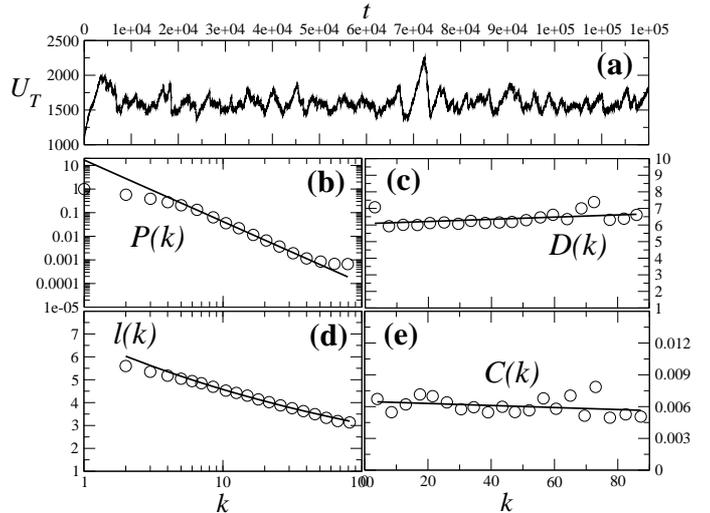}}
\vspace*{8pt}
\caption{\protect
         The network evolution of economical trades in the model
         of Ref.~\cite{CruzLind1} (see text).
         {\bf (a)} The observable $U_T$ together with the averaged
         main topological properties, all of them as functions
         of the number of trades $k$ (neighbors, degree), namely:
         {\bf (b)} The degree distribution $P(k)$,
         {\bf (c)} the degree-degree correlations $D(k)$,
         {\bf (d)} the average shortest path length $l(k)$ and
         {\bf (e)} the clustering coefficient $C(k)$.}
\label{fig4New}
\end{figure}
%%%%%%%%%%%%%%%%%%%%%%%%%%%%%%%%%%%%%%%%%%%%%%%%%%%%%%%%%%%%%%%%%%%

Since the energy consumption is necessary to establish economic 
connections, %since a single 
agents cannot leverage themselves to infinity.
To take this fact into account, each agent in the model
is not allowed to exceed its internal energy more than a 
maximum fraction of its total amount of trades.
%, i.e.~one agent cannot 
%consume more than a certain amount of labor from other agents.
%Thus, we define a quotient $d_i$ of agent $i$, which gives
%the fraction between its internal energy 
%$U_i=\beta ( k _{i,\mathrm{out}} - k _{i,\mathrm{in}})$ and its corresponding 
%``importance'' in the network, i.e.~the total amount of trading 
%connections $T_i=\beta ( k _{i,\mathrm{out}} + k _{i,\mathrm{in}})$. This quoti%ent
%has a threshold $d_{\mathrm{th}}$ below which the agent breaks its economic 
%connections:
%\begin{equation}
%d _{i}= \frac{U _{i}}{T _{i}} < d _{\mathrm{th}} .
%\label{eq:defice}
%\end{equation} 
We can regard this maximum fraction
%non-linear threshold 
as a limit for default, as in credit risk modeling \cite{MertonVasicek}. 
For the particular case of financial systems, we recently showed how 
unexpected results may appear when varying such threshold\cite{CruzLind2}.

Since trades are only regulated by supply and demand principles, the system
may lead one agent to eventually exceed its maximum fraction of total 
amount of trades.
This also occurs in real economic systems in what one calls typically
a case of insolvency:
the agent is no longer able to guarantee the ``payment'' $E_{ij}$ of the
corresponding labor units $W_{ij}$ it consumes. 
Therefore it stops to consume such labor units,
i.e.~%
%When the Eq.~(\ref{eq:defice}) is not fulfilled, 
the agent loses its 
incoming connections (consumption links).
%, i.e.~it is not able to pay to its neighbors 
%anymore and consequently they stop ``working'' for him.
%When one agent loses all its incoming connections, it collapses.

By losing its consumption links, the collapsing agent leads to the breaking 
of some production links of its neighbors, changing also their 
consumed fraction of total amount of trades, eventually also exceeding
the maximum fraction, and thus also leading to their collapse.
Consequently,
%$d_i$ eventually below $d_{\mathrm{th}}$.
%Thus, 
each collapse is able to trigger a chain reaction originating a 
branching process as illustrated in Fig.~\ref{fig2}.
In the economic context, one chain reaction is called
a ``crisis'' whereas in the physical context its is usually called 
``avalanche'', borrowing the illustrative examples of avalanches in the 
field of granular materials and complexity\cite{nuno}. 

The probability for an avalanche to involve exactly $r$
agents is given by Otter's theorem \cite{Otter,Harris,Athreya,lauritsen96}, 
which yields $P(r) \propto r ^{-\frac{3}{2}}$.
This probability is observed as long as the underlying topology 
enables a branching process in a critical state\cite{Harris}.

Still, the number of collapsed agents in a real network
is difficult to recount for. 
What is measured when an avalanche occurs in such a real
network is the number of links destroyed during the avalanche.
This number of links accounts for a macroscopic property of the system,
namely the overall product $U_T$ which sums up all outgoing product 
of all agents.
Therefore, we want to express $P(r)$ in terms of the total number of 
destroyed links.

To that end we recall the third fundamental assumption listed in the
introduction above, which considers heterogeneity among agents, where
agents tend to establish trades with those agents having already large
number of trades, i.e.~neighbors. 
Some authors in Economics call this phenomenon imitation \cite{Bouchaud}.
Physicists call it {\it preferential attachment}
and it was introduced by Simon \cite{Simon} and developed later
by Barab\'asi \cite{Barabasi_Albert_1999} in several other contexts.

The important consequence of this assumption is that the distribution of 
the number $k$ of connections one agent has follows a power-law
\begin{equation}
P(k)\sim k^{-\gamma} .
\label{eq:powerlaw}
\end{equation}

Implementing the above ingredients altogether yields a network of trades
which evolves in time. Between two time-steps, new connections and agents
are introduced
following the preferential attachment scheme and old connections are
partially removed, through the collapse of some agents, both in such 
a balance that the overall product fluctuates, as shown in
Fig.~\ref{fig4New}a. 
The increase of $U_T$, corresponds to the inflow of agents and connections
-- here, one at each time-step -- while the decrease of $U_T$ corresponds 
to a chain of collapses.

Simultaneously,
at each time-step, the underlying network topology
also ``fluctuates'' around a ``mean'' structure. In Figs.~\ref{fig4New}b-e we
show four main properties for characterizing the structure of the
network of trades, averaged over time:
the degree distribution $P(k)$, the degree-degree correlation $D(k)$
giving the average degree of neighbors of agents with $k$ neighbors,
the average shortest path length $l(k)$ and the clustering coefficient
$C(k)$\cite{Barabasi_Albert_1999}.

It is interesting to observe that while the degree distribution follows
a power-law decay at least in the middle-range of its $k$-spectrum 
(Fig.~\ref{fig4New}b), there are almost no correlations 
(Fig.~\ref{fig4New}c), with 
$D(k)\equiv \sum_{k^{\prime}} k^{\prime}P(k^{\prime}\vert k)$, typically between $6$ 
and $7$. In this context, the model of Ref.~\cite{CruzLind1} for economical
trading systems seems to belong to a general class of uncorrelated 
networks previously modeled\cite{preSatorras}.
Finally, the clustering coefficient is also approximately constant in 
the full observed $k$-spectrum, similar to other models\cite{preSatorras} 
and for the average shortest path length 
one observes a logarithmic decay, $l(k)\sim \log{k}$.

Important for our derivation in the next paragraph are the correlations,
that from Fig.~\ref{fig4New}c can be neglected. Since the full structure
shows no significant correlations, the correlations between the degree
of collapsing agents in one single chain can also be neglected.

We should emphasize that the study of distribution of destroyed 
connections is done over a probability space where the events  
occur in the agents. But the only property of an economic system 
that can be measured are the destroyed (created, for positive avalanches) 
 connections. Thus, the exercise is: taken Eq.~(\ref{eq:powerlaw}) as 
an approximation of the degree distribution $P(k)$, as in Fig.~\ref{fig4New}b; 
taken a total of $r$ affected agents, what is the amount of destroyed 
connections measured? In other words, we have a single random variable 
corresponding to the total number of affected agents, but the available measure 
of the events is 
a derived property given a realization of the random variable.

Assuming  a total number of destroyed connections $K_T$ and referencing $j*$ 
as an index over the degrees of the agents, then 
$K_T=r \sum _{j*}{k_{j*} ^{-\gamma+1}}$. If all affected agents have the 
same degree $k_{l*}$ then $K_T \propto r k_{l*}^{-\gamma+1} \propto r K_T^{-\gamma+1}$ 
and $r \propto K_T^{\gamma}$. If the affected nodes have different degrees, 
the relation of proportionality between $r$ and $K_T ^{\gamma}$ must hold 
because $r$ is fixed for the exercise so the measured property cannot grow 
with $K_T$ for one distribution of degrees $k_{j*}$ and differently for $k_{l*}$.
The proportionality coefficient can be different, but not on another function $K_T$.

Thus, to a realization $r$ of the random variable that represents the total 
number of agents collapsed, we have a measure the total destroyed connections 
proportional to $r^{1/\gamma}$. And, finally, a probability density of the 
random variable $r$, $p(r) \propto r^{-3/2}$ corresponds to a probability 
density of measure of $p(K_T) \propto K_T^{-\frac{3}{2} \gamma}$.

%\xxr{%
%To express $P(k)$ in terms of the number of destroyed connections,
%we take Eq.~(\ref{eq:powerlaw}) as an approximation of the degree 
%distribution $P(k)$ in Fig.~\ref{fig4New}b.
%Assuming such distribution together with the negligible correlations,}
%
%the number of agents with $k _j$ 
%connections involved in one avalanche with $r$ agents is 
%$n _j=r k_j^{-\gamma}$, corresponding to $k_j n _j=r k_j^{-\gamma+1}$ destroyed 
%links in agents with $k_j$ links. 
%The amount $k_j$ is a fraction of the total number
%$K_T$ of agents, say $k_j=\alpha_jK_T$. Thus, 
%$r \propto K_T^{\gamma}$, i.e.~$P(K_T) \propto K_T^{-\frac{3}{2}\gamma}$.
%

So, the fraction of avalanches of size $K_T$, i.e. involving a number 
$K_T$ of lost connections, larger than $s$ is given by
\begin{equation}
P(K_T\geq s)\propto \int _{s} ^{+\infty} {x ^{-\frac{3}{2}\gamma} dx} 
            \propto s ^{-\frac{3}{2}\gamma+1} \equiv s^{-m} .
\label{eq:avalanchefinalacum}
\end{equation}    
Equation (\ref{eq:avalanchefinalacum}) should hold for all types of 
trades and macroscopic observables of the economic product in one 
economic network, having a scale-free topology, which results from
the third fundamental assumption presented above in the Introduction.

Equation (\ref{eq:avalanchefinalacum}) establishes the relation between
the structure of microscopic interactions (connections or trades)
between agents and the distribution of the returns of one macroscopic 
quantity. The former is characterized by exponent $\gamma$ of the
degree distribution, while the latter is characterized by exponent
$m$, which according to Eq.~(\ref{eq:avalanchefinalacum}) satisfies
\begin{equation}
m=\tfrac{3}{2}\gamma-1 .
\label{m}
\end{equation}
Figure \ref{fig3}b shows explicitly this relation for each stock market
index, comparing it with the agent model that yields $m\sim 5/2$ and also
the bounding values $m_{min}$ and $m_{max}$ corresponding to the
bounding values $\gamma\in [2,3]$.

The bridge between a network of trades and macroscopic observables
is subtle: since the value of a stock index is the result of an
aggregation of successive trades -- buys and sells -- between single
agents, it can be in general taken as a variable described by the
evolution of $U_T$. In other words, the dropping and growth
periods of $U_T$, taken e.g.~as a stock index, 
are result of the underlying respective addition and removal
of connections between economical agents. 
In this context, one can draw conclusions for the evolution
of stock indices by looking the behavior of avalanches on such
networks of trades.
Next, we will derive the $\gamma$
bounding values.

%%%%%%%%%%%%%%%%%%%%%%%%%%%%%%%%%%%%%%%%%%%%%%%%%%%%%%%%%%%%%%%%
\section{Bounding values for the avalanche size distribution}
\label{distributionbounderies}   
    
All indices in Fig.~\ref{fig3}b take values around the model
prediction $m=5/2$, see Ref.~\cite{CruzLind1}, and lay within
the range $m_{\mathrm{min}}\equiv 2 < m < \tfrac{7}{2} \equiv 
m_{\mathrm{max}}$. 
From Eq.~(\ref{m}), one concludes that the above range of $m$-values 
corresponds to the range $2 < \gamma < 3$, which is a typical
range of exponent values observed in empirical scale-free 
networks specially in the economic ones like 
airports\cite{Li}, Internet\cite{barabasi} and international trade 
of products and goods\cite{Baskaran}.
With such observations we ask:
What are then the topological causes underlying the emergence
of those bounding values? In this section, we derive $m_{\mathrm{min}}$ 
and $m_{\mathrm{max}}$, applying renormalization methods\cite{Song_2005} 
to the case of undirected networks.

The bounding values of $m$ result directly from bounded 
values of $\gamma$ (see Eq.~(\ref{m})), and this latter
values can be derived under the assumption that the 
degree distribution if scale invariant.
%%%%%%%%%%%%%%%%%%%%%%%%%%%%%%%%%%%%%%%%%%%%%%%%%%%%%%%%%%%%%%%%%%%%%%%%%%%%%
\begin{figure}[htb] 
\centerline{\includegraphics[width=0.5\textwidth]{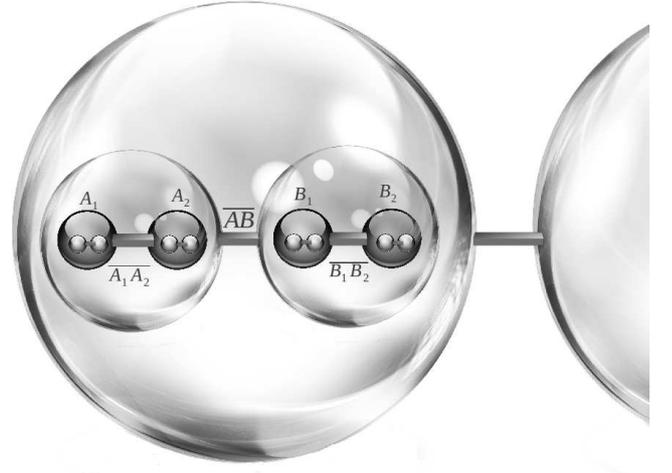}}
\vspace*{8pt}
\caption{\protect
     Illustration of renormalization in complex networks. 
     Starting at connection $\bar{AB}$ between two clusters of agents,
     one scales down finding each cluster composed by two sets of agents,
     $A_1$ and $A_2$ on the left and $B_1$ and $B_2$ on the right connected
     again by $\bar{A_1A_2}$ and $\bar{B_1B_2}$ respectively. Each 
     set of agents of this new scale can also be decomposed in 
     two connected sets and so on downscale.
     For undirected networks, the number of probable states 
     grows with $N_p^2$, with $N_p$ being the renormalized number
     of agents.}
\label{fig4}
%\end{wrapfigure}
\end{figure}
%%%%%%%%%%%%%%%%%%%%%%%%%%%%%%%%%%%%%%%%%%%%%%%%%%%%%%%%%%%%%%%%%%%%%%%%%%%%%
%%%%%%%%%%%%%%%%%%%%%%%%%%%%%%%%%%%%%%%%%%%%%%%%%%%%%%%%%%%%%%%%%%%%%%%%%%%%%
\begin{figure}[htb]
\centerline{\includegraphics[width=0.5\textwidth]{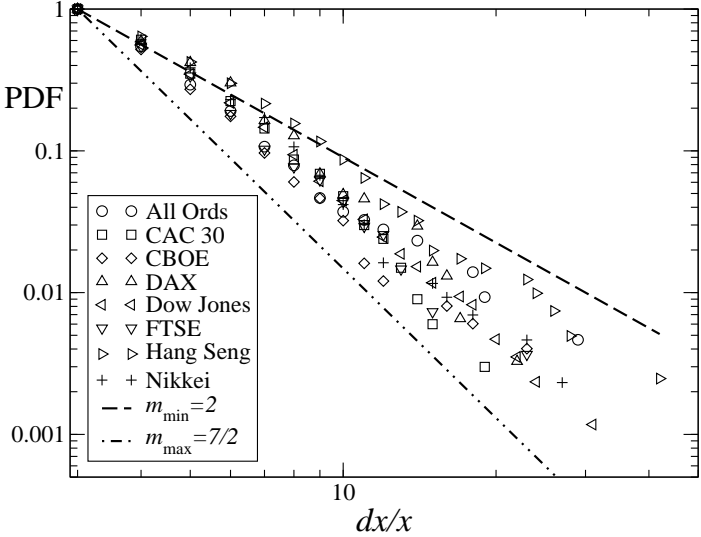}}
\vspace*{8pt}
\caption{\protect
     The bounding heavy-tails with $m_{min}=2$ and $m_{max}=7/2$ delimiting
     the return distributions of the stock indices in Fig.~\ref{fig1}a.}
\label{fig5}
%\end{wrapfigure}
\end{figure}
%%%%%%%%%%%%%%%%%%%%%%%%%%%%%%%%%%%%%%%%%%%%%%%%%%%%%%%%%%%%%%%%%%%%%%%%%%%%%

As mentioned above, links are either outgoing or incoming and the 
probability for an agent to have $k$ outgoing links -- or correspondingly 
$k$ incoming links -- depends on the scale one is considering:
at each scale $p$ there is a fraction $P(p,k)$ of agents with $k$ connections.
Figure \ref{fig4} illustrates three successive scales $p=1,2$ and $3$ for
directed networks.
When the connections are directed, from one scale to the next there are $N$
admissible connections\cite{Song_2005,Song_2006}.
Undirected networks can be regarded as compositions of two directed 
networks, since the degree law $P(p,k)$ is the 
probability for a agent to have $k$ start links or $k$ end links
{\it indistinctly}.
Consequently, when going from one scale to the next, the renormalization
generates $N^2$ admissible states leading to 
$\frac{dN^2 P(p,k)}{dp} = 0$.

Therefore, the self-similar transformation of the agent degree, i.e.~the number 
of links in a agent will be ruled by 
\begin{equation}
N^2 P(k) dk = N^2_{p} P(k_{p}) dk_{p}
 \label{eq:agentdegreerenormdirec}
\end{equation}
where $k$ and $k_{p}$ symbolize, respectively, the total and 
renormalized number of links and $N$ and $N_{p}$ are the correspondent 
number of agents. 
  
The power-law in Eq.~(\ref{eq:powerlaw}) is invariant under 
renormalization, i.e.~$P(k)\sim k^{-\gamma}$ and $P(k _{p})\sim k_p^{-\gamma}$.
Defining $l_p$ as the distance between agents at a given scale $p$, as
the average number of links separating a randomly chosen pair of agents at 
a given scale $p$, the fractal dimension $d_B$ of the network can then be 
calculated using the box-counting technique \cite{falconer,Song_2005}:
\begin{equation}
N_{p} = N l_p^{-d_B}   .
 \label{eq:agentboxcounting}
\end{equation}
Similarly, the number of links scale as
\begin{equation}
k_{p} = k l_p^{-d_{k}}  .
 \label{eq:linkboxcounting}
\end{equation}
And finally, substituting Eqs.~(\ref{eq:powerlaw}), (\ref{eq:agentboxcounting}) 
and (\ref{eq:linkboxcounting}) in Eq.~(\ref{eq:agentdegreerenormdirec})
yields
\begin{equation}
\gamma = 1+2\frac{d_B}{d_{k}} .
 \label{eq:gammafinal}
\end{equation}

This result retrieves a topological constraint for the value of $\gamma$ 
and, consequently, for the ``weight'' of the heavy-tail in the degree 
distribution. 
It is known that whenever the above results holds the corresponding
degree distribution is invariante under renormalization (see Supplementary Material
of Ref.~\cite{Song_2006}).
At each scale the number of connections of each agent varies between two
limit cases, one where each agent connects to only one neighbor, and another
where everybody is connected with everybody else, within the same scale.
In the first limit case, each agent links to a single neighbor at each scale 
and consequently the connections will scale like the agents, $d_k=d_B$, yielding 
$\gamma=3$. 
In the other limit case, agents should connect to all neighbors at a each scale,
i.e.~for each set of $N_p$ agents we find $N_p^{2}-N_p\sim N_p^2$ links
and thus $d_{k}=2 d_{B}$, i.e.~$\gamma=2$.
Since both limit cases yield a relation between $d_k$ and $d_B$,
the same conclusion should hold even in non-fractal networks 
similar to what is reported in Ref.~\cite{Song_2006}.
Not all power-law degree distributions are invariant under renormalization. Still, it is
reasonable to expect that even in the case they are not strictly invariant, the exponent 
characterizing their power-law degree distribution should lie between the two limit cases of 
(invariant) degree distributions. It is therefore a general result as shown next for
observed financial indices.

In a real network, at each scale each agent should have a typical
number of connections between these two extremes, namely one and
$N$, resulting in an exponent $\gamma$ between $2$ and $3$ and
in the corresponding two bounding values for $m$ in Eq.~(\ref{m}),
$m_{min}=2$ and $m_{max}=7/2$.
Figure \ref{fig5} plots the cumulative distributions for all indices
together with the boundaries $P_{min}(dx/x)\sim (dx/x)^{-m_{min}}$  and
$P_{max}(dx/x)\sim (dx/x)^{-m_{max}}$.
This is an important result since, independently of the network 
complexity, the return distributions are characterized by heavy-tails 
in a limited range of frequencies.
Consequently, the amplitude of the associated risk measure is also 
limited.

To finish this section we discuss possible applications
from these findings.
Having such bounding values an important application deals
with risk evaluation.
The ability to measure risk is fundamental when we talk about any economic 
activity. When for instance banks lend people money to buy houses,
%or investment houses buy companies capital to make money with 
%their development, 
they must have a way of estimating the risk of those activities. 
In short, what is the most one can lose on a particular investment?
%what is the minimum loss incurred in an investment in $A$\%
%worse cases in a given time horizon[WHICH REF HERE???]. 
The financial property Value at Risk, or simply $VaR$, provides an answer.
%This measure is called the Value-at-Risk($VaR_{\alpha}(T)$) with an $\alpha$ 
%degree of confidence in a $T$ time horizon and it is the conceptual network 
%that is the base for most of risk regulation all around.
Value at Risk evaluates the percentile of the predictive probability 
distribution for the size of a future financial loss. 
Mathematically, for a prescribe $\alpha$ degree of confidence and within a 
time horizon $\Delta t$ the value at risk is defined as the value $x^{\ast}$
such that
\begin{equation}
VaR_{\alpha}(x^{\ast},\Delta t)\equiv \int_{x^{\ast}}^{\infty} p(x)dx = 1-\alpha
\label{var}
\end{equation}
where $p(x)$ is the PDF for the loss or the
negative return of a economic variable relevant for the 
intended investment. If one is dealing with shares portfolio, the relevant 
variable would be one
%given macroeconomical index
such as the ones in Fig.~\ref{fig3}, symbolized here as $x$.
%To measure $VaR_{\alpha}(T)$ we take a sample of $n$ returns of a price $X$ 
%after a $T$ interval of time, making $(\frac{X(t+T)-X(t)}{X(t)}$. 
%With that we find the $A$\% worse cases, 
%$x^{(\alpha)}(X)= sup {x|P[ X \leq x] \leq \alpha}$  and 
%$VaR_{\alpha}(T)=-x^{(\alpha)}(X)$. 
Value-at-Risk framework was created by J.P.~Morgan\cite{Riskmetrics} based on 
Gaussian returns but the framework is valid for other more general
%under the assumption of other 
continuous distributions of returns.

The confidence of the estimate given for $VaR_{\alpha}(x^{\ast},\Delta t)$
depends therefore on the choice of the PDF $p(x)$ for the losses or returns.
Since under the assumptions above we can take the PDF $p(x)$ as a Pareto
distribution and since we can bound the exponent value defining such a Pareto
distribution, we get a straightforward way for bounding any estimate of $VaR$.
While $VaR$ is a measure of risk, i.e.~a risk model for 
estimating how much one can lose in a specific investment based in some 
functional form of $p(x)$, by bounding its value through the two bounding
exponent values deduced above we are able to evaluate how ``risky'' are
such risk models and risk measures. By ``risky'' we refer more specifically
to the choice of $p(x)$ when evaluating the risk measure, in this case 
$VaR_{\alpha}$.
In other words, our bounding exponent values can provide us with a way to
evaluate the ``model risk'' of a particular model for risk evaluation.

\section{Conclusion: towards a risk model}
\label{conclusion}            

In this paper we derive a relation that bridges between basic economic 
principles and features of the empirical return distributions, namely
the values of the exponent describing their heavy-tails.
First, we show that three fundamental assumptions in economics\cite{Lipsey} 
suffice to explain the emergence of economic links according to
a power-law and relate the exponent of the heavy-tails observed in 
the macroscopic variable with the exponent of the degree distribution
describing the underlying topology of the network.
Second, we show that this latter exponent for the agents degree 
distribution assumes values in the range between two and three, which
combined with the first results delimits the heavy-tails of return
distribution between $2$ and $7/2$.

These findings help to solve the controversy about
Mandelbrot hypothesis \cite{Mandelbrot2} 
that the distribution of financial returns are explained
by L\'evy distributions, and therefore would yield 
an exponent smaller than $3$.
Some authors have argued against Mandelbrot hypothesis,
basing their positions \cite{Borland_Bouchaud2005} in empirical measures
of the return distributions which yield
exponents larger than $3$.
In this paper we presented evidence that both statements are in fact
correct. Each one is considering a different effect of the same phenomenon: 
though what we measure in the time series of returns are links 
between agents, corresponding to exponents larger than $3$,
the random variables behind L\'evy distributions are the 
agents, which corresponding to the exponent $\gamma$ which is
limited by $2<\gamma<3$.

Finally, since the exponent is bounded, the total risk associated 
with the process being observed is also bounded between one lower 
and one upper boundary values,
enabling one to actually measure the ``risk'' of 
a particular model for risk evaluation. We described the particular
case of risk measure, namely the Value at Risk, but other approaches
could be also taken, for example the expected 
shortfall\cite{AcerbiTasche,Acerbi}, which considers the average Value at Risk
with respect to its confidence level $\alpha$.

\section*{Acknowledgments}
The authors thank Nuno Ara\'ujo for helpful discussions and
also PEst-OE/FIS/UI0618/2011 for partial financial support.
PGL thanks 
{\it Funda\c{c}\~ao para a Ci\^encia e a Tecnologia – Ci\^encia 2007} for
financial support.

%\section*{References}
\bibliographystyle{elsart-num}
\bibliography{boundfin}
%%%%%%%%%%%%%%%%%%%%%%%%%END BIBLIOGRAPH%%%%%%%%%%%%%%%%%%%%%%%%%%%%%

\end{document}